\documentclass[twocolumn,aps,pre,groupedaddress, amsmath,amssymb]{revtex4}

\usepackage{graphicx}% Include figure files
\usepackage{dcolumn}% Align table columns on decimal point
\usepackage{bm}% bold math
\usepackage{epstopdf}% Allow to use eps image format.
\usepackage{color}
\usepackage{pdfpages}

\begin{document}

\title{Shear-banding and superdiffusivity in entangled polymer solutions}
\author{Seunghwan Shin}
\affiliation{Department of Chemical Engineering and Materials Science, University of Minnesota, Minneapolis, Minnesota 55455, USA}
\author{Kevin D. Dorfman}
 \email{dorfman@umn.edu}
\affiliation{Department of Chemical Engineering and Materials Science, University of Minnesota, Minneapolis, Minnesota 55455, USA}
\author{Xiang Cheng}
 \email{xcheng@umn.edu}
 \affiliation{Department of Chemical Engineering and Materials Science, University of Minnesota, Minneapolis, Minnesota 55455, USA}
\date{\today}

\begin{abstract}

Using high-resolution confocal rheometry, we study the shear profiles of well-entangled DNA solutions under large amplitude oscillatory shear (LAOS) in a rectilinear planar shear cell. With increasing Weissenberg number (Wi), we observe successive transitions from normal Newtonian linear shear profiles to wall-slip dominant shear profiles and finally to shear-banding profiles at high Wi. To investigate the microscopic origin of the observed shear banding, we study the dynamics of micron-sized tracers embedded in DNA solutions. Surprisingly, tracer particles in the shear frame exhibit transient super-diffusivity and strong dynamic heterogeneity. The probability distribution functions of particle displacements follow a power-law scaling at large displacements, indicating a L\'{e}vy-walk-type motion, reminiscent of tracer dynamics in entangled wormlike micelle solutions and sheared colloidal glasses. We further characterize the length and time scales associated with the abnormal dynamics of tracer particles. We hypothesize that the unusual particle dynamics arise from localized shear-induced chain disentanglement.              

\end{abstract}

\maketitle

\section{Introduction}
Shear banding, namely the coexistence of multiple shear rates in a sheared sample, has been observed in numerous complex fluids such as worm-like micelle solutions \cite{Walker01,Callaghan08}, concentrated colloidal suspensions \cite{Cohen06,Koumakis12}, low-density attractive colloidal gels \cite{Grenard14}, dense assemblies of soft microgels \cite{Divoux10} and granular materials \cite{Cheng06}. Due to its fundamental importance in understanding the response of materials under large and fast shear deformations, shear banding has been a persistent focus of rheological studies \cite{Olmsted08,Schall10,Divoux16,Fielding16}. Although entangled linear polymers were predicted to display shear banding over 30 years ago based on the then newly developed Doi-Edwards theory \cite{McLeish86}, the existence of shear banding in entangled linear polymers remains under debate. Experimental evidence supporting the existence of shear banding in entangled linear polymers was first reported by Wang and coworkers under both steady and time-dependent shear conditions \cite{Tapadia06a,Ravindranath08a,Ravindranath08,Boukany08a,Boukany09a,Boukany09,Wang11,Cheng12}. However, subsequent studies suggested that the observed banding flows may arise from the edge fracture in cone-plate rheometers and/or experimental artifacts \cite{Hu10,Ravindranath12,Hu12,Li13,Wang14,Li14}. Although the controversy has not been fully settled \cite{Hayes08,Hayes10,Noirez09,Jaradat12}, these experimental findings have aroused intensive numerical and theoretical investigations that attempt to rationalize the existence of shear banding \cite{Fielding03,Cao12,Cromer13,Cromer14,Mohagheghi15,Mohagheghi16a,Mohagheghi16b,Moorcroft13}. Mechanisms including shear-induced concentration fluctuations \cite{Fielding03,Cromer13,Cromer14}, localized chain disentanglements \cite{Mohagheghi15,Mohagheghi16a,Mohagheghi16b} and transient instabilities triggered by stress overshoot in shear startup \cite{Moorcroft13,Cao12,Mohagheghi16b} have been proposed for the possible existence of shear banding in entangled polymeric fluids.           

In spite of the theoretical progress, important experimental questions remain open, largely contributing to the on-going controversy. First, can one observe shear banding in polymeric fluids after ruling out possible experimental defects? If so, what are the microscopic dynamics of polymer shear-banding flows? Here, we address these two fundamental questions by using a high-resolution rheo-optical apparatus designed to avoid experimental artifacts \cite{Hayes08,Hayes10}. Our experiments directly verify the existence of shear banding in highly entangled polymer solutions under planar large amplitude oscillatory shear (LAOS). Moreover, we find that the dynamics of polymer solutions in the two co-existing bands differ substantially. In the high-shear-rate band, embedded tracer particles exhibit a L\'{e}vy-walk type of motion and a transient super-diffusive behavior with a high degree of dynamic heterogeneity. Long temporal correlations but short-range spatial correlations are observed for the abnormal dynamics of tracer particles. Such unusual tracer dynamics provide important insights into the origin of the shear-banding flows in polymeric fluids and suggest localized shear-induced chain disentanglement. 

\section{Materials and methods}

\subsection{Sample preparation and characterization}

Our experiments use calf thymus DNA (double-stranded, 75 kbp, 4.9$\times 10^{9}$ g/mol, Affymetrix) as our model polymer. Solutions of calf thymus DNA were prepared in an aqueous $2\times$ TBE buffer (180 mM Tris base, 180 mM Boric acid, 5.6 mM EDTA). The buffer strongly screens electrostatic interactions, so that DNA molecules behave similarly to neutral semi-flexible polymers \cite{Tree13}. The moderate pH ($\approx8$) maintained by the buffer also prevents the denaturation of DNA. We added fluorescently-tagged, carboxy-terminated polystyrene (PS) micro-beads (radius $a=0.55$ $\mu$m, Thermo Fisher Scientific) in the DNA solution as tracer particles. The volume fraction of PS spheres is kept so low ($< 0.03\%$) that the interaction between tracers can be safely ignored. 

The final concentration of DNA was confirmed using a UV/Vis spectrometer (NanoDrop, Thermo Scientific). In the shear-banding experiments, the concentration of DNA was fixed at $8.3\pm0.4$ mg/mL. Based on a previous study using $\lambda$-DNA \cite{Teixeira07}, the overlap concentration of $\lambda$-DNA in a high ionic strength buffer is $c^{*}=0.064$ mg/mL. As $c^{*}\approx 3M/(4\pi R_{g}^{3}N_{A}) \sim M^{1-3\nu}$, where $M$ is the molecular weight, $R_g$ is the radius of gyration, $N_A$ is the Avogadro number and $\nu=3/5$, we estimated $c^{*}=0.051$ mg/mL for our samples of calf thymus DNA. Thus, the DNA concentration used in our experiments is two orders magnitude larger than $c^{*}$, ensuring a high degree of chain entanglement. 

\begin{figure}
\centering
\includegraphics[width=0.45\textwidth]{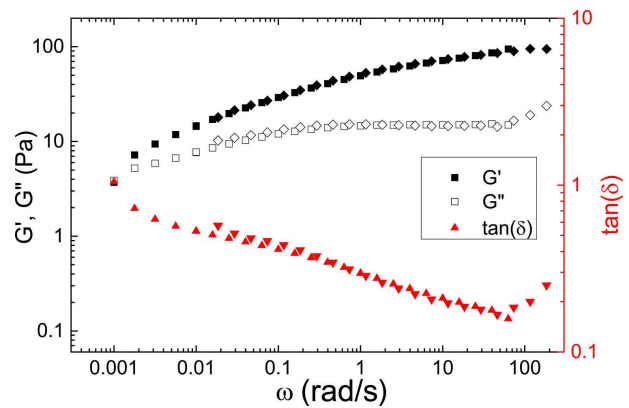}

\caption{Linear viscoelasticity of a highly-entangled DNA solution. The storage modulus ($G'$), the loss modulus ($G''$) and the loss tangent ($\tan\delta$) were measured in a frequency sweep under SAOS. Solid squares and diamonds are for $G'$. Empty squares and diamonds are for $G''$. Solid triangles (up-pointing and down-pointing) are for $\tan\delta$. The results were used to determine the relevant tube parameters $\tau_{d}$, $\tau_{R}$, $G_{N}^{0}$ and $Z$. Squares and up-pointing triangles were obtained at the room temperature 23 $^\circ$C, whereas diamonds and down-pointing triangles were taken at 13 $^\circ$C and shifted to the room temperature using the principle of time-temperature superposition.  }
\label{fig:Figure1}
\end{figure}

We estimated the tube parameters from the linear viscoelasticity of the solution. Frequency sweep experiments under small amplitude oscillatory shear (SAOS) were conducted using a commercial cone-plate rheometer (AR-G2, TA Instruments) (Fig.~\ref{fig:Figure1}). We used time-temperature superposition to obtain the final result. We chose $G'$ at the frequency where the loss tangent $\tan\delta$ is minimal as the plateau modulus, $G_N^{0} \approx 100$ Pa. The average number of entanglements per chain ($Z$) was calculated by Ferry's temporary network formula \cite{Ferry80}, 
\begin{equation}
Z=\frac{5}{4}\frac{M}{M_{e}}=\frac{5}{4}G_{N}^{0}(cRT)^{-1} \approx 300,
\end{equation} 
where $c = 8.3$ mg/mL. Thus, the mesh size of the polymer network is about 90 nm based on $Z$. Reptation time was given by the reciprocal overlap frequency when $G'=G''$, $\tau_{d} \approx 900$ s. Finally, the Rouse relaxation time is $\tau_{R}=\tau_{d}/(3Z) \approx 1$ s \cite{Larson03}. Thanks to the long contour length of DNA molecules, we can achieve a high entanglement at a relatively low DNA concentration so that the elastic modulus of the solution is sufficiently small for us to load and shear the sample \cite{Boukany08a,Boukany09a,Boukany09}. However, due to the slow relaxation at such a high DNA concentration, we could not reach the terminal relaxation regime even at the lowest frequency of our rheometer.

Since the non-uniformity of DNA concentration in a solution may trigger shear banding flows \cite{Fielding03,Cromer13}, we test the uniformity of our high-concentration DNA solutions in three different ways. First, we take a small volume ($\sim$ 2 $\mu$L) of DNA solution from a concentrated sample at five different, well-separated positions in a sample vial. UV/Vis measurements are then made at least five times for the solution taken at each position. The variation of the measured concentrations is within $5\%$ of the average concentration, indicating that both the device measurement error and the spatial concentration variation are reasonably small in our experiments. Second, we also measure the velocity profiles at low Weissenberg number (Wi) ($f=0.1$ Hz). Presumably, if a sample is locally heterogeneous, we would observe non-uniform shear deformations of the sample at low Wi. Furthermore, the diffusion of DNA molecules due to the concentration gradient would result in the change of shear profiles over time. We find that the velocity profiles of our samples are always stable and linear, again indicating a good uniformity of the solutions. Lastly, sample homogeneity can also be qualitatively checked from the distribution of PS tracer particles. As mentioned above, we seed a small amount of fluorescent PS tracers in DNA solutions for flow visualization. After slowly stirring the solution for mixing, any aggregated particle clusters in the original solution are well dispersed. The uniformity of the particle distribution is directly confirmed from microscope images, which again suggests the uniformity of DNA concentrations after stirring.
       
\subsection{Experimental setup}

\begin{figure}
	\centering
	\includegraphics[width=0.37\textwidth]{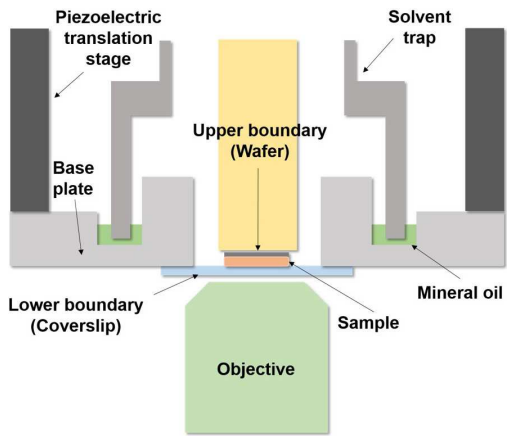}
	
	\caption{Schematic of the experimental setup. The area of the wafer is $5 \times 5$ mm$^2$. The thickness of the sample is 100 $\mu$m. While the wafer is fixed, the coverslip is moved by the piezoelectric stage, which induces a linear shear in the confined sample. The solvent trap creates a sealed chamber, preventing the evaporation of the solvent. An inverted confocal microscope (represented here by the objective) is used for imaging the shear flow within the sample.}
	\label{fig:Figure2}
\end{figure}

Figure~\ref{fig:Figure2} shows the schematic of our experimental setup. A small amount of the highly-entangled DNA solution (20 $\mu$L) is loaded in a custom shear cell. The shear cell has a rectilinear plane-plane geometry with a square top plate of length $L=5$ mm made of an etched silicon wafer and a much larger bottom plate made of a normal microscope coverslip. While the top plate is kept stationary, the bottom plate is driven by a piezoelectric actuator, which exerts a sinusoidal oscillatory shear in the sample confined between the two plates following $\gamma_0(t)=(A_0/H)\sin(2\pi f t)$ (Supplementary Video 1). Here, $A_0$ is the applied shear amplitude, $f$ is the applied shear frequency, and $H$ is the gap size between the top and bottom plates. In our experiments, we keep $H \le 100$ $\mu$m such that the aspect ratio of the shear cell is large with $L/H \ge 50$. The detailed description of the setup can be found in Refs.~\cite{Cheng11,Lin14}. The corresponding Weissenberg number based on the Rouse relaxation time is Wi = $\left( A_0/H \right)$De, where De $=2\pi \tau_R f$ is the Deborah number of large amplitude oscillatory shear (LAOS). The shear cell is coupled to a fast inverted spinning-disk confocal microscope for high-resolution flow field visualization \cite{Cheng11,Lin14}. 

Before each of our measurements, we allow the sample to equilibrate in the shear cell without shear for at least 30 minutes, about twice as long as $\tau_{d}$. An oscillatory shear is then applied at the desired shear frequency and strain amplitude for another 30 minutes. The preshear allows the sample reach a steady state before measurements. Videos are then taken at different heights above the moving plate and analyzed with Particle Image Velocimetry (PIV) (Supplementary Video 1). The confocal microscope scans two-dimensional images at 200 frames per second. These images have an area of $100 \times 100$ $\mu$m$^2$ when a $60\times$ lens was used, and $300 \times 300$ $\mu$m$^2$ when a $20\times$ lens was used, which typically contain a few hundred tracer particles in the field of view. An image sequence containing two to four shear cycles is taken at each height $y$ before the focal plane is adjusted a different $y$ either manually or automatically. The time for adjusting the focal plane is irrelevant, since the shear velocity at each height is analyzed separately and our sample has reached the steady state before measurements. At each height, the peak velocity of the oscillatory shear is determined by applying Fast Fourier Transformation (FFT) to the velocity as a function of time. We typically measure velocities at 10 different heights to construct one shear profile. Three shear profiles are measured for the sample at a given shear condition. The final shear profile is an average of these three independent experimental runs. Error bars reflect the variation of the three measurements. 

Our setup shares the same advantage as the confocal-rheoscopic device introduced in Ref.~\cite{Boukany15}, where high-resolution imaging of velocity profiles within narrow gaps was achieved. However, Ref.~\cite{Boukany15} applies steady shear to entangled polybutadiene solutions in a rotational rheometer. Here, we used a planar Couette cell, which eliminates stress inhomogeneity and streamline curvature that may potentially trigger shear banding \cite{Divoux16}. Indeed, the rectilinear shear cell with a large aspect ratio $L/H$ provides an ideal geometry excluding or significantly reducing potential experimental artifacts \cite{Hu10,Li13} such as the curvature of material lines, secondary flows and thermal and edge effects \cite{Hayes08,Hayes10}.

Lastly, we discuss the influence of the shear boundary. (1) For the vertical shear boundary, in all the experiments reported in this article, the bottom shear plate is made of a normal microscope coverslip. The top shear plate is constructed from a silicon wafer.  The wafer is roughened by reactive-ion etching, where we use 40 sccm CF$_4$ and 4 sccm O$_2$ at pressure 100 mtorr and power 100 W. The etching time is slightly less than one minute. From atomic force microscopy (AFM), the roughness of the wafer surface is $34.5 \pm 1.7$ nm, calculated based on the root-mean-squared heights. Symmetric shear boundaries with both the top and bottom plates made of glass coverslips have also been tested, which yield qualitatively similar behavior (results not shown). (2) For the lateral boundary condition, the perimeter of the cell is left open. Samples overflow outside the shear cell laterally and are held fixed by the capillary force at the contact line with the bottom plate. Since the confined gap of the shear cell is much smaller than the lateral dimension, we do not expect the lateral boundary condition strongly affects the flow profile of samples at the center of the cell \cite{Hayes08,Hayes10}, where we conduct all our measurements. Rheological measurements have indeed shown that the influence of the lateral boundary on the shear profile only extends into the sample on the order of the gap thickness \cite{Reddy13}, which is far away from the center of the cell. A similar geometry has been adopted before in studying the shear profile of polymer samples without shear banding \cite{Hayes08,Hayes10}. It is worth noting that although the measurements on velocity profiles at the center of the cell are not affected by the lateral boundary, rheological measurement of the stresses may be affected depending on the boundary condition. This is because stress measurements average all the samples in the cell including those near the lateral boundary. To further test the influence of the edge, we measure the shear profiles at different gap thicknesses in the small gap limit for semi-dilute DNA solutions. The profiles are all linear as expected without any perceptible influence from the lateral boundary.

\section{Results}

\subsection{Shear-banding of entangled DNA solutions}
To probe the response of highly entangled DNA solutions under large amplitude oscillatory shear (LAOS), we vary the gap thickness of our shear cell, $H$, and the shear frequency, $f$, while keeping the strain amplitude fixed at $\gamma_{0}=1.5$. 

\begin{figure}
	\begin{center}
		\includegraphics[width=2.45 in]{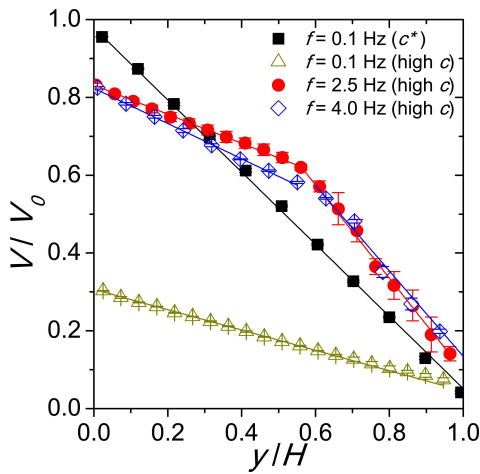}
		%select pdftexify command to run jpg or pdf files
	\end{center}
	\caption[Shear response of DNA solution]{Normalized velocity profiles of DNA solutions under shear. Black squares are for a semi-dilute DNA solution at the overlap concentration $c^* = 0.05$ mg/mL. Others are for the high concentration, entangled DNA solution. Solid lines indicate piecewise linear fits. A Cartesian coordinate system is defined, where $x$, $y$ and $z$ are the flow, shear gradient and vorticity directions, respectively. Velocity amplitudes, $V(y)$, are normalized by the velocity amplitude of the moving plate, $V_0 = 2\pi f A_0$ with $A_0= 150$ $\mu$m. } 
	\label{fig:Figure3}
\end{figure}   

Figure~\ref{fig:Figure3} illustrates typical behaviors of the DNA solution under different shear conditions. For comparison, we also show the usual linear velocity profile of a semi-dilute DNA solution at the overlap concentration. For the entangled DNA solution at the high concentration, the sample shows a strong, asymmetric wall slip near the moving bottom plate at low $f$ (low Wi). A stronger slip occurs at the moving bottom plate of a smooth glass coverslip. In contrast, the slip at the stationary top plate made of a roughened silicon wafer is much weaker. The applied strain is relaxed dominantly through the local yielding of a thin layer of DNA molecules near the moving plate. As a result, the local shear strain in the bulk of the material $\gamma$ is much smaller than $\gamma_{0}$. The wall-slip dominant linear velocity profile eventually turns into clear shear-banding profiles at higher $f$ (higher Wi), which show small but comparable wall slips at both plates. This trend is qualitatively similar to the results from the Wang group \cite{Wang11}. In addition, we also observe the similar trend when using monodisperse $\lambda$-DNA (data not shown). As such, our experiments with a high-resolution confocal rheoscope directly demonstrate the existence of shear banding in entangled DNA solutions in rectilinear shear under LAOS.

\begin{figure}
	\centering
	\includegraphics[width=0.37\textwidth]{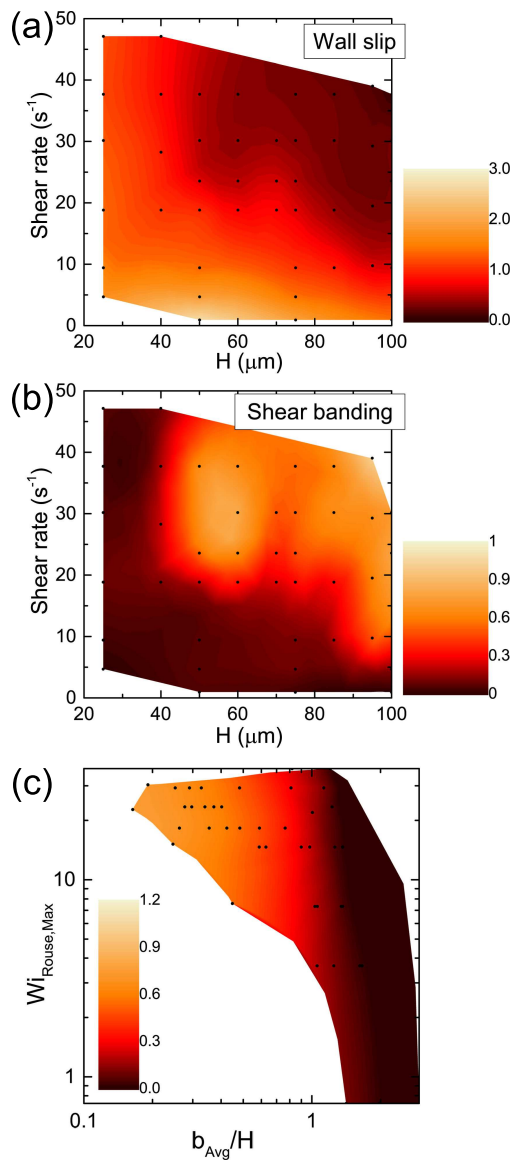}
	
	\caption{Phase diagram of highly-entangled calf thymus DNA solutions under LAOS. The amplitude of shear strain is fixed at $\gamma_{0}=1.5$. (a) Contour map of the wall-slip parameter, $(b_{h}+b_{l})/2H$, at different shear rates $\dot{\gamma}_{0}$ and gap thicknesses $H$. Solid dots correspond to our experimental points. (b) Contour map of the shear-banding parameters, $\vert\dot{\gamma}_{h}-\dot{\gamma}_{l}\vert/\dot{\gamma}_{0}$, from the same experiments. (c) Contour map of the shear-banding parameter in terms of Wi and the slip parameter, $b_{avg}/H = (b_{h}+b_{l})/2H$.}
	\label{fig:Figure4}
\end{figure}

To quantify the shape of the velocity profiles, we fit the profiles piecewisely with linear lines (Fig.~\ref{fig:Figure3}). A banding parameter can then be defined as the difference between two shear rates relative to the applied shear rate, $\vert\dot{\gamma}_{h}-\dot{\gamma}_{l}\vert/\dot{\gamma}_{0}$, where $\dot{\gamma}_{h,l}$ are the shear rates of the high and low shear-rate bands, respectively. $\dot{\gamma}_{0} = V_0/H =2\pi f\gamma_0$ is the applied shear rate, where $V_0$ is the velocity amplitude of the moving plate. Furthermore, to characterize the degree of wall slip, we measure the slip lengths of a velocity profile at both the stationary and moving plates by extrapolating the profile to the non-slip velocities at 0 and $V_0$. The normalized average slip length is then calculated using $(b_{h}+b_{l})/2H$, where $b_{h,l}$ are the slip lengths at the stationary and moving plates, respectively.

Based on the banding parameter and the slip length, we map a detailed phase diagram for the shear response of entangled DNA solutions under LAOS (Fig.~\ref{fig:Figure4}). Fig.~\ref{fig:Figure4} shows that shear banding flows are more favorable when the degree of wall slip is small. With increasing $\dot{\gamma}_{0}$ and $H$, linear profiles with strong wall slips gradually give away to shear-banding flows. Shear banding and wall slip are complementary to each other. Following the previous study by Wang and co-workers \cite{Wang11}, we also plot the banding parameter in terms of Wi and the slip parameter in Fig.~\ref{fig:Figure4}c. The result again demonstrates that shear banding sets in when wall-slip is suppressed and Wi is large.

\subsection{Super-diffusivity of tracer particles}

\begin{figure}
\begin{center}
\includegraphics[width=2.65 in]{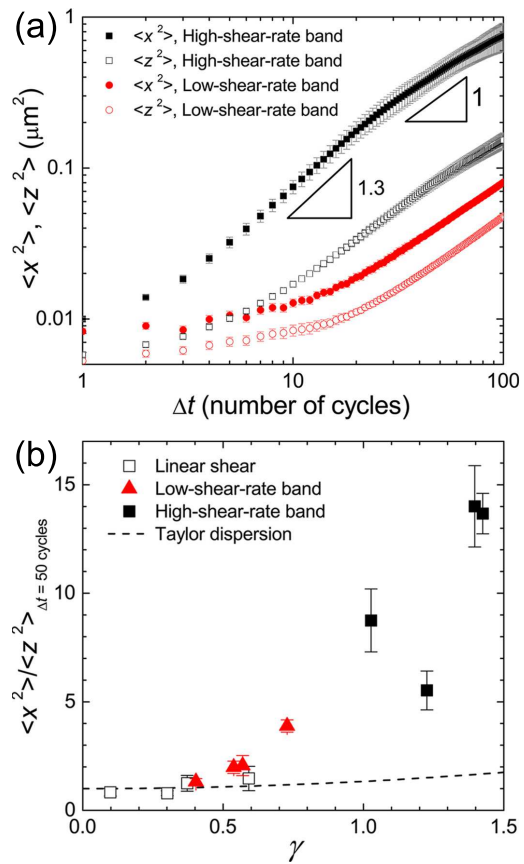}
%select pdftexify command to run jpg or pdf files
\end{center}
\caption[Particle dynamics in the shear frame]{Dynamics of particles in the shear frame. (a) Mean-squared displacements (MSDs) of particles along the flow ($x$) and vorticity ($z$) directions in the two co-existing shear bands. The slopes indicate the superdiffusive motion of particles at intermediate times and the diffusive motion at long times. (b) The ratio of MSDs, $\langle x^2 \rangle/\langle z^2 \rangle$, at $\Delta t=50$ cycles versus local shear strains, $\gamma$. The dashed line indicates the prediction of Taylor dispersion. Different $\gamma$ are achieved by varying $A_0$ while keeping $f = 4$ Hz.} 
\label{fig:Figure5}
\end{figure}

To explore the microscopic dynamics of the shear banding flow, we shall focus on the behavior of the solution at high Wi with fixed $f=4$ Hz (De $\approx 25$) and $A_0=120$ $\mu$m (Wi $\approx$ 30). In the lab frame, the average sinusoidal motion of tracer particles dominates (Supplementary Video 1), allowing us to accurately extract the velocity profiles (Fig.~\ref{fig:Figure3}). In order to reveal the fluctuation of particle dynamics around the mean shear flow, we acquire images of the tracer particles in the shear frame at a frequency equal to $f$, so that a sequence of strobed images of the tracer particles are taken at the same phase of successive shear cycles (Supplementary Videos 2 and 3). For imaging convenience, we choose the phase at $\pi/2$ where $\gamma_0(t)$ is largest. Fig.~\ref{fig:Figure5}a shows the mean-squared displacements (MSDs) of particles in the shear frame in the two co-existing shear bands, respectively. Particles diffuse faster in the high-shear-rate band close to the stationary plate, indicating a lower effective shear viscosity in that band, consistent with the overall velocity profile (Fig.~\ref{fig:Figure3}). In both the low- and high-shear-rate bands, a strong anisotropy in particle dynamics is observed. Particles exhibit faster dynamics along the direction of the shear flow. Quantitatively, we plot the ratio of MSDs along the flow ($x$) and vorticity ($z$) directions, $\langle x^2 \rangle/\langle z^2 \rangle$, at $\Delta t = 50$ cycles under different shear conditions, where particles show clear long-time diffusions (Fig.~\ref{fig:Figure5}b). $\langle x^2 \rangle/\langle z^2 \rangle$ increases with the amplitude of local shear strains $\gamma$. At low Wi, when the velocity profile is linear, the ratio follows the prediction of the classic Taylor dispersion with $\langle x^2 \rangle/\langle z^2 \rangle = 1 + \gamma^2/3$ \cite{Leahy13}. However, the prediction fails in the shear banding flow, especially in the high-shear-rate band, where particle dynamics in $x$ are more strongly enhanced. 

More interestingly, particles in the high-shear-rate band show a transient superdiffusive motion when $\Delta t=3$ to 20 cycles (Fig.~\ref{fig:Figure5}a), where $\langle x^2 \rangle \sim \Delta t^\alpha$ with $\alpha = 1.41 \pm 0.13$ obtained from an average of 25 independent experimental runs. Similar transient superdiffusivity has also been observed in stationary worm-like micelle solutions \cite{Ott90} and in sheared dense colloidal suspensions \cite{Koumakis12,Zausch08}. Indeed, one can easily identify the particles with anomalous fast motions in Supplementary Video 3, taken in the high-shear-rate band. A fraction of the particles jump over a large distance along $x$ within a short time interval, $\Delta t$, whereas other particles hardly diffuse within the same $\Delta t$. In sharp contrast, no such jumpers can be observed in the co-exiting low-shear-rate band close to the moving plate (Supplementary Video 2). We also observe a small fraction of jumpers next to the walls when there is a strong wall slip.

\begin{figure}
\begin{center}
\includegraphics[width=2.8 in]{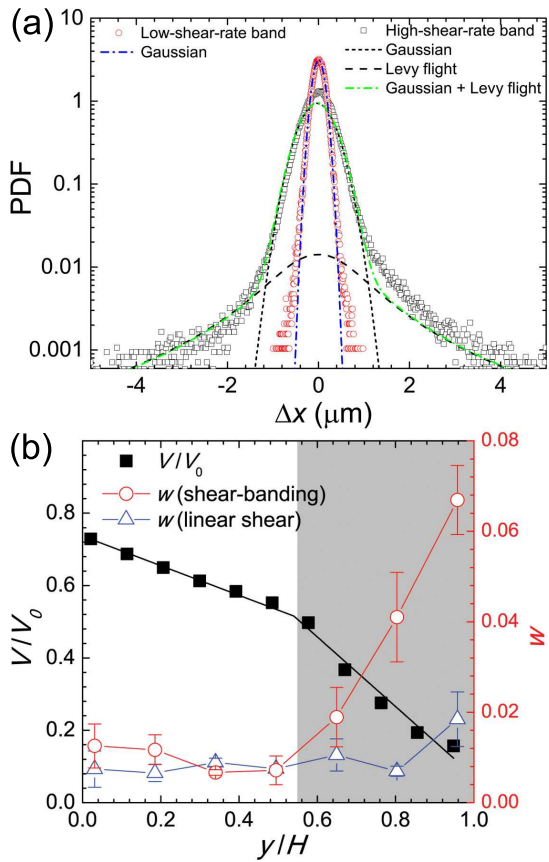}
%select pdftexify command to run jpg or pdf files
\end{center}
\caption[PDF and jumper fractions]{L\'{e}vy walk of tracer particles. (a) Probability distribution functions (PDFs) of particle displacements over 20 shearing cycles in the high- and low-shear-rate bands. PDF in the low-shear-rate band is fitted by a Gaussian distribution. PDF in the high-shear-rate band is fitted by Eq.~\ref{equation1}. (b) The fraction of jumpers performing the L\'{e}vy walk, $w$, at different heights. The corresponding velocity profile is shown for comparison. The shaded area indicates the high-shear-rate band. The fraction of jumpers in the sample with a linear velocity profile at $A_0 = 37.5$ $\mu$m (Wi $\approx 9.4$) is also shown.} 
\label{fig:Figure6}
\end{figure} 

We further quantify the anomalous particle dynamics by plotting the probability distribution functions (PDFs) of the displacements of particles along the flow direction, $\Delta x$, over $\Delta t = 20$ cycles (Fig.~\ref{fig:Figure6}a). While PDF of particles in the low-shear-rate band can be well described by a Gaussian distribution, the jumpers in the high-shear-rate band give rise to long power-law tails in the PDF at large $\Delta x$, indicating a L\'{e}vy walk of the jumpers \cite{Ott90,Klafter11}. Such an observation inspires us to fit the PDFs using 
\begin{equation}
\text{PDF}(\Delta x) = \frac{(1-w)}{\sqrt{2\pi\sigma^2}}e^{-\frac{\Delta x^2}{2\sigma^2}} + \frac{w(1+\mu)\sin \left(\frac{\pi}{1+\mu}\right)}{2\pi\left(1+|\Delta x|^{1+\mu}\right)},
\label{equation1}
\end{equation}                     
where the first term gives the Gaussian core at small $\Delta x$ with a standard deviation $\sigma$ and the second term gives the distribution of the L\'{e}vy walk with PDF$(\Delta x)\sim \Delta x^{-(1+\mu)}$ at large $\Delta x$. The power-law exponent of the distribution $1 \le \mu < 2$ is linked to the anomalous diffusion exponent MSD $\sim t^{3-\mu}$ \cite{Klafter11}. We find $\mu = 1.45 \pm 0.17$ in the PDF, agreeing with the MSD measurements. The weighting factor, $w(y)\in[0,1]$, gives the relative contribution of the two distributions and, therefore, can be used to quantify the fraction of jumpers. We plot $w(y)$ in both the linear and nonlinear shear flows (Fig.~\ref{fig:Figure6}b). Jumpers are not observed in the linear shear flow except near the wall with wall-slip. Although the fraction of jumpers is also low in the low-shear-rate band of the shear-banding flow, it increases sharply in the high-shear-rate band, consistent with our direct observation.

\begin{figure}
\begin{center}
\includegraphics[width=2.55 in]{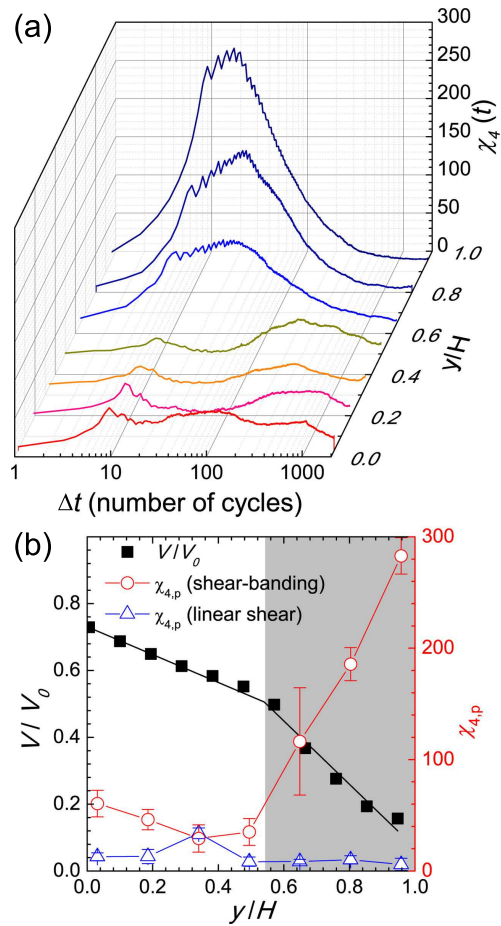}
%select pdftexify command to run jpg or pdf files
\end{center}
\caption[Dynamic heterogeneity]{Dynamic heterogeneity of tracer particles. (a) Four-point susceptibility, $\chi_4(t)$, at different heights in the shear-banding flow. (b) The peak value of $\chi_4(t)$, $\chi_{4,p}$, at different heights. The corresponding shear-banding velocity profile is shown for comparison. The shaded area indicates the high-shear-rate band. $\chi_{4,p}$ for the sample with a linear shear profile at $A_0=50$ $\mu$m (Wi $\approx 12.5$) is also shown.} \label{fig:Figure7}
\end{figure}

The co-existence of fast jumpers and slow diffusive particles reveals a strong dynamic heterogeneity in the sheared solution. We quantify this dynamic heterogeneity by measuring a four-point susceptibility $\chi_4(t) = N\left[\langle q_s(t)^2\rangle - \langle q_s(t)\rangle^2\right]$, where $q_s(t)$ is the self-overlap function of particles \cite{Berthier11}: $q_s(t)=(1/N)\sum_{i=1}^{N}w\left(|x_i(t)-x_i(0)|\right)$ with $w=1(0)$ if $|x_i(t)-x_i(0)|<(>)a$. Here, $x_i(t)$ is the position of the tracer particle $i$ at time $t$ and $N$ is the total number of tracer particles. $\chi_4(t)$ measures the standard deviation of particle dynamics quantified by $q_s(t)$; a larger $\chi_4$ corresponds to a stronger dynamic heterogeneity. $\chi_4(t)$ at different heights in the shear-banding flow are shown in Fig.~\ref{fig:Figure7}a. Strong dynamic heterogeneities emerge in the high-shear-rate band at the intermediate time scales, coincident with the superdiffusive regime in the MSDs (Fig.~\ref{fig:Figure7}a). The well-defined peak of $\chi_4(t)$ is reminiscent of particle dynamics in crowded environments near the glass transition \cite{Berthier11}. Quantitatively, we compare the peak value of $\chi_4$, $\chi_{4,p}$, in different shear bands (Fig.~\ref{fig:Figure7}b). It is clear that strong dynamic heterogeneities only develop in the high-shear rate band of the shear-banding flow, consistent with the presence of the jumpers in Fig.~\ref{fig:Figure6}b. Note that the heterogeneous dynamics at micron scales strongly limit the possibility of a secondary flow triggered by the edge failure as the origin of the anomalous particle dynamics and, for that matter, the shear banding in our LAOS experiments \cite{Hu10,Li13}. Unless microscopic vortices of tens of microns---the average spacing between jumpers and diffusers---were formed in the secondary flow, the flow would lead to uniform particle drift at this small scale. 

\subsection{Correlation of jumper dynamics}

\begin{figure}
\begin{center}
\includegraphics[width=3 in]{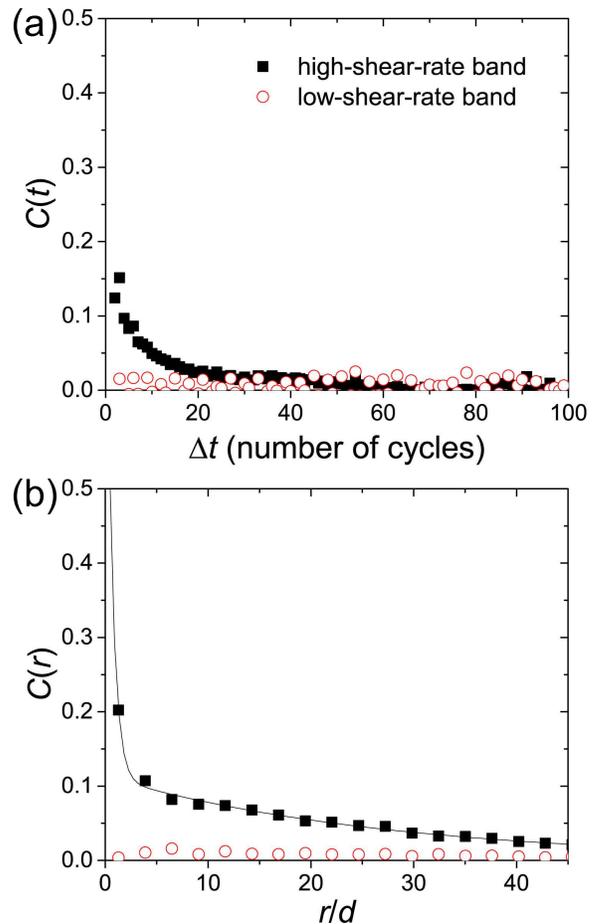}
%select pdftexify command to run jpg or pdf files
\end{center}
\caption[jumper dynamics correlation]{Correlation of jumper dynamics. (a) The autocorrelation of particle velocities, $C(t)$, in the low- and high-shear-rate band. Time is measured in the unit of the number of shear cycles. (b) The spatial correlation of particle velocities, $C(r)$, in the low- and high-shear-rate band. The radial distance $r$ is normalized by the diameter of tracer particles $d = 1.1$ $\mu$m. The solid line indicates a double-exponential fit.} 
\label{fig:Figure8} 
\end{figure} 

The dynamics of jumpers in the high-shear-rate band exhibit interesting spatiotemporal features. Some jumpers sustain slow drifts over multiple cycles after performing a large leap and finally turning back into normal diffusion (Supplementary Video 3). Moreover, neighboring jumpers may show strong correlated motions. To extract the associated time and length scales of jumper dynamics, we measure the temporal and spatial correlations of jumper dynamics. The results reveal crucial information on the underlying dynamics of the sheared entangled polymer network. 

We obtain the trajectories of jumpers using Particle Tracking Velocimetry (PTV), where we define a particle as a jumper when its maximal displacement over one cycle within the entire trajectory is larger than one particle diameter. The autocorrelation of jumper dynamics, $C(t)$, can be written as, 
\begin{equation}
C(t) = \frac{\langle \bm{v_{i}}(t_{0}) \cdot \bm{v_{i}}(t_{0}+t) \rangle_{i,t_0}}{\langle \bm{v_{i}}(t_{0}) \cdot \bm{v_{i}}(t_{0})\rangle_{i,t_0}},
\label{equation3}
\end{equation}  
where $\bm{v_{i}}(t_{0})$ is the velocity of particle $i$ at time $t_{0}$ given by $\lbrack \bm{r_{i}} (t_{0}+1)-\bm{r_{i}} (t_{0}) \rbrack$ and $t$ is the time difference in terms of the number of cycles. The average is taken over all particles $i = 1,N$ and all $t_0$. For comparison, we also calculate $C(t)$ for particles in the low-shear-rate band. Whereas particles in the low-shear-rate band show no temporal correlation as expected for normal diffusion (Fig.~\ref{fig:Figure8}a), jumpers in the high-shear-rate band show a weak but long correlation with a time constant $\tau = 11.2 \pm 1.7$ cycles. 

We also quantify the spatial correlation of jumper dynamics. We find the spatial correlation is isotropic in the flow-vorticity plane. Thus, we define the spatial correlation, $C(r)$, based on the radial variable $r$ alone as
\begin{equation}
C(r) = \frac{\langle \bm{v}(\bm{r_{i}},t_{0}) \cdot \bm{v}(\bm{r_{j}},t_{0})\rangle_{i,j,t_0}}{\langle \bm{v}(\bm{r_{i}},t_{0}) \cdot \bm{v}(\bm{r_{i}},t_{0})\rangle_{i,t_0}},
\label{equation4}
\end{equation}
where $i$ and $j$ are indices denoting tracer particles and $r=\vert \bm{r_{i}} - \bm{r_{j}} \vert$. The average is taken over all the particle pairs and the initial time $t_0$. Similar to the autocorrelation, diffusers in the low-shear-rate band show almost zero spatial correlation as expected (Fig.~\ref{fig:Figure8}b). The spatial correlation of jumpers in the high-shear-rate bands shows a more interesting behavior. With increasing $r$, a strong short-range correlation is followed by a much weaker but long-range correlation. Accordingly, we fit $C(r)$ using a double-exponential decay $C(r) = A\exp(-r/l_s) + (1-A)\exp(-r/l_l)$. The strong short-range correlation gives a length scale $l_s = 0.81 \pm 0.23$ $\mu$m, which is about 0.75 particle diameters or 10 mesh sizes, whereas the weak long-range correlation shows a length scale $l_l = 30.2 \pm 2.6$ $\mu$m, comparable to the width of the high-shear-rate band. The existence of two different spatial correlation length scales can be observed in Supplementary Video 3. 

\section{Discussion}

The superdiffusivity and dynamic heterogeneity of tracer particles provide important insights into the microscopic polymer dynamics under shear. We hypothesize that the abrupt motion of jumpers arises from localized shear-induced chain disentanglement and the breakage of local entangled polymer network. The motion of tracer particles is strongly constrained by the local entangled mesh of DNA chains. If the shear deformation is affine and the dense mesh of chains that regulates the tracer motion is intact, the positions of tracers with respect to the local entangled network should remain the same (assuming the diffusion of tracers is negligible). This explains our observations in the low-shear-rate band, where particles barely move. We argue that the observed abnormal dynamics of jumpers in the high-shear-rate band is possible only if the mesh network around the region containing jumpers experiences entanglement failure and deforms in a non-affine manner. The event releases the shear stress built up around jumpers induced by the non-affine deformation of the network around the particles, which pushes the particles out of the entangled ``cage'', i.e., the local constraints imposed by the entangled polymer network on the particles. Although the associated length scale $l_s = 0.82$ $\mu$m is 10 times larger than the mesh size, it is small enough to safely exclude the influence of global effects such as the secondary flow on the correlation of particle motions \cite{Hu10,Li13}. Nevertheless, it is not clear if the long-range correlation at the scale of $l_l$ is induced by the secondary flow. If this long-range correlation indeed arises from the secondary flow, why does the secondary flow concentrate only in the high-shear-rate band? Hence, our hypothesized ``localized disentanglement'' should be understood on the scale of $l_s$. Within this hypothesis, the polymer solution should have a lower degree of entanglement in the high-shear-rate band and, therefore, exhibit a lower viscosity, a picture consistent with a recent numerical finding \cite{Mohagheghi15}. Moreover, the presence of jumpers and the small $\chi_4$ peak near the shear plates with strong wall slip (Figs.~\ref{fig:Figure6}b and \ref{fig:Figure7}b) indicate shear-induced chain disentanglement near the boundary, consistent with previous theories on wall slip, where apparent slip occurs due to the disentanglement of polymer chains near the wall \cite{Hatzikiriakos15}. This observation suggests a common origin of shear banding and wall slip \cite{Divoux16, Wang11}.

Independent of the validity of the hypothesis, our results on unusual tracer dynamics provide a solid experimental benchmark for testing microscopic theories of the rheology of concentrated polymeric fluids. Our results also pose further questions and directions. First, it is important to theoretically understand the time and length scales of jumper dynamics and, particularly, the two different length scales of the spatial correlation of jumper dynamics. Second, it is not clear why the fraction of jumpers and the dynamic heterogeneity increase continuously with height, rather than plateauing, in the high-shear-rate band (Figs.~\ref{fig:Figure6}b and \ref{fig:Figure7}b). Last, it is also interesting to verify if shear banding exists in steady shear flows \cite{Ravindranath08} and measure the corresponding shear-induced particle dynamics. For LAOS at small De approaching the limit of steady shear, our sample shows a strong wall slip (Fig.~\ref{fig:Figure4}), preventing us from testing the possible steady-state shear banding. Future experiments with rough shear plates and higher solvent viscosity may help to reduce the degree of wall slip and allow us to probe this interesting regime.

\section{Conclusions}

We studied the response of concentrated DNA solutions under LAOS using a custom high-resolution rectilinear confocal shear cell. The setup allows us to avoid potential experimental artifacts and experimentally verify the existence of shear-banding in highly entangled polymer solutions under LAOS. Moreover, we found that the embedded tracer particles in the shear-banding polymeric fluids display markedly different dynamics in the two co-existing shear bands. While particles in the low-shear-rate bend show normal diffusive motions, particles in the high-shear-rate band exhibit transient super-diffusivity, obeying the distribution of L{\'e}vy walks. A high degree of dynamic heterogeneity were also observed exclusively in the high-shear-rate band. We quantitatively analyzed the associated length and time scales of abnormal particle dynamics. We hypothesized that these dynamic features are associated with shear-induced chain disentanglement and breakage of local entanglement network in the high-shear-rate band, although it is not clear if such a microscopic picture can quantitatively predict the length and time scales of tracer particles. This poses a theoretical challenge for the future work.     

\begin{acknowledgments}

We thank F. Bates, C. Macosko, C. Schroeder and L. Walker for fruitful discussions and P. Agrawal for preparing the etched silicon wafer. S. S. acknowledges financial support from the Kwanjeong Educational Foundation. The work was supported by NSF CBET-1700771. Portions of this work were performed in UMN Nanofabrication Center, which receives a partial support from NSF National Nanotechnology Infrastructure Network (NNIN), and UMN College of Science and Engineering Polymer Characterization Facility, which receives capital equipment funding from NSF through the UMN MRSEC program (DMR-1420013).   
\end{acknowledgments}

\bibliographystyle{apsrev4-1}

\begin{thebibliography}{99}


\bibitem{Walker01} L. M. Walker, Curr. Opin. Colloid Interface Sci. {\bf 6}, 451 (2001).   
\bibitem{Callaghan08} P. T. Callaghan, Rheol. Acta {\bf 47} 243 (2008). 
\bibitem{Cohen06} I. Cohen, B. Davidovitch, A. B. Schofield, M. P. Brenner, and D. A. Weitz, Phys. Rev. Lett. {\bf 97}, 215502 (2006).
\bibitem{Koumakis12} N. Koumakis, M. Laurati, S. U. Egelhaaf, J. F. Brady, and G. Petekidis, Phys. Rev. Lett. {\bf 108}, 098303 (2012). 
\bibitem{Grenard14} V. Grenard, T. Divoux, N. Taberlet, and S. Manneville, Soft Matter {\bf 10}, 1555 (2014).

\bibitem{Divoux10} T. Divoux, D. Tamarii, C. Barentin, and S. Manneville, Phys. Rev. Lett. {\bf 104}, 208301 (2010).
\bibitem{Cheng06} X. Cheng, J. B. Lechman, A. Fernandez-Barbero, G. S. Grest, H. M. Jaeger, G. S. Karczmar, M. E. M{\"o}bius, and S. R. Nagel, Phys. Rev. Lett. {\bf 96}, 038001 (2006).
\bibitem{Olmsted08} P. D. Olmsted, Rheol. Acta {\bf 47}, 283 (2008).
\bibitem{Schall10} P. Schall and M. van Hecke, Annu. Rev. Fluid Mech. {\bf 42}, 67 (2010).
\bibitem{Divoux16} T. Divoux, M. A. Fardin, S. Manneville, and S. Lerouge, Annu. Rev. Fluid Mech. {\bf 48}, 81 (2016). 
\bibitem{Fielding16} S. M. Fielding, J. Rheol. {\bf 60}, 821 (2016).

\bibitem{McLeish86} T. McLeish and R. Ball, J. Polym. Sci. Pol. Phys. {\bf 24}, 1735 (1986).
\bibitem{Tapadia06a} P. Tapadia, S. Ravindranath, and S.-Q. Wang, Phys. Rev. Lett. {\bf 96}, 196001 (2006).
\bibitem{Ravindranath08a} S. Ravindranath and S.-Q. Wang, J. Rheol. {\bf 52}, 341 (2008). 
\bibitem{Ravindranath08} S. Ravindranath, S.-Q. Wang, M. Olechnowicz, and R. P. Quirk, Macromolecules {\bf 41}, 2663 (2008).
\bibitem{Boukany08a} P. E. Boukany, Y. T. Hu, and S.-Q. Wang, Macromolecules {\bf 41}, 2644 (2008).  

\bibitem{Boukany09a} P. E. Boukany and S.-Q. Wang, J. Rheol. {\bf 53}, 73 (2009).
\bibitem{Boukany09} P. E. Boukany and S.-Q. Wang, Soft Matter {\bf 5}, 780 (2009).
\bibitem{Wang11} S.-Q. Wang, S. Ravindranath, and P. E. Boukany, Macromolecules {\bf 44}, 183 (2011).
\bibitem{Cheng12} S. Cheng and S.-Q. Wang, J. Rheol. {\bf 56}, 1413 (2012).
\bibitem{Hu10} Y. T. Hu, J. Rheol. {\bf 54}, 1307 (2010).

\bibitem{Ravindranath12} S. Ravindranath, Y. Wang, P. E. Boukany, and X. Li, J. Rheol. {\bf 56}, 675 (2012).
\bibitem{Hu12} Y. T. Hu, J. Rheol. {\bf 56}, 683 (2012).
\bibitem{Li13} Y. Li, M. Hu, G. B. McKenna, C. J. Dimitriou, G. H. McKinley, R. M. Mick, D. C. Venerus, and L. A. Archer, J. Rheol. {\bf 57}, 1411 (2013).
\bibitem{Wang14} S.-Q. Wang, G. Liu, S. Cheng, P. E. Boukany, Y. Wang, and X. Li, J. Rheol. {\bf 58}, 1059 (2014).
\bibitem{Li14} Y. Li, M. Hu, G. B. McKenna, C. J. Dimitriou, G. H. McKinley, R. M. Mick, D. C. Venerus, and L. A. Archer, J. Rheol. {\bf 58}, 1071 (2014).

\bibitem{Hayes08} K. A. Hayes, M. R. Buckley, I. Cohen, and L. A. Archer, Phys. Rev. Lett. {\bf 101}, 218301 (2008). 
\bibitem{Hayes10} K. A. Hayes, M. R. Buckley, H. Qi, I. Cohen, and L. A. Archer, Macromolecules {\bf 43}, 4412 (2010). 
\bibitem{Noirez09} L. Noirez, H. Mendil-Jakani, and P. Baroni, Macromol. Rapid Comm. {\bf 30}, 1709 (2009).
\bibitem{Jaradat12} S. Jaradat, M. Harvey, and T. A. Waigh, Soft Matter {\bf 8}, 11677 (2012).  
\bibitem{Fielding03} S. M. Fielding and P. D. Olmsted, Phys. Rev. Lett. {\bf 90}, 224501 (2003).

\bibitem{Cao12} J. Cao and A. E. Likhtman, Phys. Rev. Lett. {\bf 108}, 028302 (2012).
\bibitem{Cromer13} M. Cromer, M. C. Villet, G. H. Fredrickson, and L. G. Leal, Phys. Fluids {\bf 25}, 051703 (2013).  
\bibitem{Cromer14} M. Cromer, G. H. Fredrickson, and L. G. Leal, Phys. Fluids {\bf 26}, 63101 (2014). 
\bibitem{Mohagheghi15} M. Mohagheghi and B. Khomami, ACS Macro Lett. {\bf 4}, 684 (2015).
\bibitem{Mohagheghi16a} M. Mohagheghi and B. Khomami, J. Rheol. {\bf 60}, 849 (2015). 

\bibitem{Mohagheghi16b} M. Mohagheghi and B. Khomami, J. Rheol. {\bf 60}, 861 (2015). 
\bibitem{Moorcroft13} R. L. Moorcroft and S. M. Fielding, Phys. Rev. Lett. {\bf 110}, 086001 (2013).
 

\bibitem{Tree13} D. R. Tree, A. Muralidhar, P. S. Doyle, and K. D. Dorfman, Macromolecules {\bf 46}, 8369 (2013).

\bibitem{Teixeira07} R. E. Teixeira, A. K. Dambal, D. H. Richter,  E. S. G. Shaqfeh, and S. Chu, Macromolecules, {\bf 40}, 2461 (2007).
\bibitem{Ferry80} J. D. Ferry, {\it Viscoelastic Properties of Polymers, 3rd ed.} (Wiley, 1980).
 
\bibitem{Larson03} R. G. Larsonn, T. Sridhar, L. G. Leal, G. H. McKinley, A. E. Likhtman, and T. C. B. McLeish, J. Rheol. {\bf 47}, 809, (2003)
%\bibitem{Supplemenetary} See Supplementary Material at XXX for details on sample %preparation and additional experimental results. 
\bibitem{Cheng11} X. Cheng, J. H. McCoy, J. N. Israelachvili, and I. Cohen, Science {\bf 333}, 1276 (2011).

\bibitem{Lin14} N. Y. C. Lin, J. McCoy, X. Cheng, B. Leahy, J. Israelachvili, and I. Cohen, Rev. Sci. Instrum. {\bf 85}, 033905 (2014).
\bibitem{Boukany15} P. E. Boukany, S.-Q. Wang, S. Ravindranath, and L. J. Lee, Soft Matter {\bf 11}, 8058 (2015).  
\bibitem{Reddy13} N. K. Reddy, C. Ruth,   and C. Clasen, Bulletin of The Society of Rheology 84th Annual Meeting (2013).
\bibitem{Leahy13} B. D. Leahy, X. Cheng, D. C. Ong, C. Liddell-Watson, and I. Cohen, Phys. Rev. Lett. {\bf 110}, 228301 (2013). 
\bibitem{Ott90} A. Ott, J. P. Bouchaud, D. Langevin, and W. Urbach, Phys. Rev. Lett. {\bf 65}, 2201 (1990). 

\bibitem{Zausch08} J. Zausch, J. Horbach, M. Laurati, S. U. Egelhaaf, J. M. Brader, T. Voigtmann and M. Fuchs, J. Phys. Condens. Matter {\bf 20}, 404210 (2008).
\bibitem{Klafter11} J. Klafter and I. M. Sokolov, {\it First Steps in Random Walks} (Oxford Univ. Press, Oxford, UK, 2011).
\bibitem{Berthier11} L. Berthier, G. Biroli, J.-P. Bouchaud, L. Cipelletti, and W. van Saarloos, {\it Dynamical Heterogeneities in Glasses, Colloids and Granular Media} (Oxford Univ. Press, Oxford, UK, 2011).
\bibitem{Groisman00} A. Groisman and V. Steinberg, Nature {\bf 405}, 53 (2000).
\bibitem{Hatzikiriakos15} S. G. Hatzikiriakos, Soft Matter {\bf 11}, 7851 (2015).


\end{thebibliography}

\end{document}